\documentclass[12pt,epsfig]{article}
\usepackage{amsmath}
\usepackage{graphicx}
\usepackage{epstopdf}
\usepackage{titlesec}
\usepackage{hyperref}
\usepackage[usenames, dvipsnames]{color}
\titleformat*{\section}{\large\bfseries\sffamily}
\titleformat*{\subsection}{\small\bfseries\sffamily}
\begin{document}
\begin{center}
\textbf {Distinction of groups of gamma-ray bursts in the BATSE catalog through fuzzy clustering}\\
\end{center}
\begin{center}
	Soumita Modak$^*$\\
	Department of Statistics, Basanti Devi College\\ affiliated with\\  University of Calcutta\\
	147B, Rash Behari Ave, Kolkata- 700029, India\\
	email: soumitamodak2013@gmail.com
	
\end{center}
\begin{abstract}
In search for the possible astrophysical sources behind origination of the diverse gamma-ray bursts, cluster analyses are performed to find homogeneous groups, which discover an intermediate group other than the conventional short and long bursts. However, very recently, few studies indicate a possibility of the existence of more than three (namely five) groups. Therefore, in this paper, fuzzy clustering is conducted on the gamma-ray bursts from the final
`Burst and Transient Source Experiment' catalog to cross-check the significance of these new groups. Meticulous study on individual bursts based on their memberships in the fuzzy clusters confirms the previously well-known three groups against the newly found five.
\end{abstract}
keywords: {Statistical machine learning; Data analysis; Gamma ray bursts; Fuzzy clustering.}
\section{Introduction}
In the search for astrophysical sources behind origination of the diverse gamma-ray bursts (GRBs), scientists have been trying to find the homogeneous classes existing in these most luminous explosions in the Universe through cluster analysis. It is an evident fact that distinct clustering  methods applied to different variables on GRBs change individual membership of the bursts which fall in the overlapping regions among the clusters. However, we attempt to reach the robust classes which can expose the inherent clustering nature of the GRBs. Early studies of Venera GRB data collected through KONUS experiment find two different groups of short and long bursts (see, Mazets et al. 1981; Norris et al. 1984). The Burst and Transient Source Experiment (BATSE) on
board the COMPTON Gamma-Ray Observatory (CGRO) operating between the years 1991 and 2000 has presently the largest data set on 2702 GRBs (Meegan et al. 1992; Fishman et al.
	1994; Meegan et al. 1996; Paciesas et al. 1999; Horv\'{a}th 2002), which are our working data. Based on the duration distribution, Kouveliotou et al. (1993) classify the bursts into short ($T_{90} < 2$ s) and long ($T_{90} > 2$ s) groups ($T_{90}$ is a duration variable
representing the time of a burst's 90\% 
flux arrival in seconds, abbreviated to s). Despite significant duration overlap between the two groups (Toth et al. 2019), the short group is hard and the long one is soft by means of their spectral hardness ratios; moreover, their physical differentiation is  supported by information on prompt-emission (Kouveliotou et al. 1993; Gehrels et al. 2009; Zhang et al. 2012), afterglows, host galaxies and redshift distributions (Zhang et al. 2009;
Berger 2011, 2014; Levan et al. 2016). Short bursts are generally believed to have compact binary mergers, like merger of two neutron stars or merger
of a neutron star with a black hole, as their progenitors (Paczy\'{n}ski 1986; Usov 1992; Bloom et al. 2006; Nakar 2007; Berger 2014); whereas massive stellar collapse is believed to generate the long bursts (Woosley
1993; Paczy\'{n}ski 1998; Woosley \& Bloom 2006; Blanchard et al.
2016). Automatic statistical cluster analyses (Horv\'{a}th 1998, 2002) hints at possibility of more than two classes with short ($T_{90}<2$ s), intermediate ($2<T_{90}<10$ s) and long ($T_{90}>10$ s) groups. These  classes are subsequently confirmed by bivariate clustering based on duration and hardness variables, where the intermediate group is found to be the softest along with conventional short--hard and long--soft groups (Horv\'{a}th et al. 2004, 2006). 
Later on, multivariate studies incorporating more variables confirm the existence of a third group; where the hard--short class of dimmer bursts with the lowest mean fluence reveal to possess an average duration of $T_{90}<2$ s, and the traditional soft--long bursts with average $T_{90}>2$ s are divided into intermediate duration faint bursts with low fluence and long duration bright bursts with high fluence (Mukherjee et al. 1998; Balastegui et al. 2001). The intermediate group is later witnessed to have a larger average duration, i.e. $10<T_{90}<30$ s (Chattopadhyay et al. 2007; Modak et al. 2018). Over the past few decades, BATSE GRBs of various sizes with differently chosen variables have been classified using efficient statistical methods: be it parametric classical model-based approaches like Gaussian-mixture, Dirichlet process (Horv\'{a}th 1998; Horv\'{a}th et al. 2006; Chattopadhyay et al. 2007), nonparametric data-mining techniques like k-means, hierarchical clustering (Mukherjee et al. 1998; Balastegui et al. 2001; Chattopadhyay et al. 2007; Modak et al. 2018), or advanced machine learning algorithms like neural network clustering, kernel principal component analysis (Balastegui et al. 2001; Modak et al. 2018), which conclude three statistically significant clusters of GRBs. 
 
In this regard, it is worth mentioning that cluster analyses of GRB data sets collected by other satellites also disclose an intermediate class supporting the presence of three groups, e.g. Beppo–SAX bursts
(Horv\'{a}th 2009), Swift/BAT data (Horv\'{a}th et al. 2008, 2010; Veres et al. 2010; Koen \& Bere 2012; Tsutsui \& Shigeyama 2014; Horv\'{a}th \& T\'{o}th 2016), 
RHESSI data set (\v{R}\'{\i}pa et al. 2009; \v{R}\'{\i}pa \& M\'{e}sz\'{a}ros 2016) and Fermi/GBM data (Horv\'{a}th et al. 2018). This intermediate group might have emerged from merger of a massive white dwarf with a neutron
	star (Chattopadhyay et al. 2007; King et al. 2007; Modak et al. 2018). However, some authors doubt that this class of GRBs might be caused by instrumental and
	sampling biases (Hakkila et al. 2000; 2003; Rajaniemi \& M\"{a}h\"{o}nen 2002); whereas some think of this group as a possible composition of the short and the long bursts which violate their predominantly believed correspondence to merger and collapse models respectively (Bromberg et al. 2013; Zitouni et al. 2015). Amid the rising concern that not all GRBs can be classified as either short bursts generated by merger model or long bursts originated by collapse model (Horv\'{a}th et al. 2018; Toth et al. 2019), astrophysical sources behind the intermediate group are yet to be confirmed.

Recently, Chattopadhyay \& Maitra (2017, 2018) study the BASTE catalog and state that there may be five clusters in the GRBs, although astrophysical distinction among the groups is not yet established. On the other hand, Toth et al. (2019) dismiss these five groups as insignificant caused just by further division of the three established classes. Also, Acuner \& Ryde (2018)
find five clusters in the Fermi/GBM catalog using model-based clustering, which after interpretation with respect to external variables boil down to two major clusters of photospheric origin and synchrotron origin. However, Chattopadhyay \& Maitra (2017) perform multivariate cluster analysis using Gaussian-mixture model and discover five distinct classes of GRBs in terms of the Bayes'
Information Criterion (BIC, Schwarz 1978). Again, in Chattopadhyay \& Maitra (2018), multivariate clustering based on $t-$mixture model results in five clusters as indicated by BIC. Nonetheless, Toth et al. (2019) point out that these new groups arise due to mere numerical separation of the peak flux variable; and they conclude that it is unlikely to have new sources behind the additional clusters caused by model-based clustering methods relying on a finite mixture of probability distributions with each distribution corresponding to a cluster. Data when fail to satisfy such parametric distributional assumptions can produce considerably misleading results. Hence, robust nonparametric methods are often opted for a better solution in analyzing the challenging astronomical data sets (Feigelson \&
Babu 2013; Bandyopadhyay \& Modak 2018; Modak \& Bandyopadhyay 2019; Modak et al. 2020). In Modak et al. (2018), $k-$means clustering (Hartigan \&  Wong 1979) based on the relevant kernel principal components (Sch\"{o}lkopf \& Smola 2002; Hofmann et al. 2008; Ishida \& Souza 2013), extracted through a novel kernel (Modak et al. 2017), rules out the possibility of five clusters provoked by noisy data and finally provides three  significant clusters of GRBs in terms of the Dunn index (Dunn 1974). 

Now, instead of exploring the clusters indicated by some automatic cluster validity index, in this study our objective is shifted to solution of the recently raised burning conflict between the five new clusters and the three well-known groups. Therefore, we explore both three and five clusters of the BATSE catalog in detail through nonparametric fuzzy clustering. Here, instead of hard clustering where each GRB is assigned to only one particular cluster, we give them membership values which explain how likely a GRB is to belong to any of the clusters. In contrast to explanation of the cluster-wise average properties using subjective expertise (Toth et al. 2019), we perform extensive investigation of individual GRBs objectively in terms of their memberships in the fuzzy clusters. We apply the efficient statistical machine learning algorithm `FANNY' (Kaufman \& Rousseeuw 2005), which does not assume any model assumption and hence can robustly reveal the inherent clusters present in the GRBs. It statistically confirms the previously well-known three groups while explaining the new five classes as insignificant division of the old three, which supports the claim of Toth et al. (2019). 

The paper is organized as follows. Section 2 outlines the data set and Section 3 discusses the methods. Section 4 explains the results with Section 5 concluding.

\section{Data set}
The current BATSE Gamma-Ray Burst Catalog\footnote{https://gammaray.nsstc.nasa.gov/batse/grb/catalog/current/} (Meegan et al. 1992; Fishman et al. 1994; Meegan et al. 1996; Paciesas et al. 1999; Horv\'{a}th 2002; Toth et al. 2019) provides information for the following observed variables of GRBs.
Fluence variables: $F_1,F_2,F_3,F_4$ are time-integrated fluences in  $20-50$, $50-100$, $100-300$ and $>300$ keV spectral channels respectively; flux variables: $P_{64},P_{256},P_{1024}$ are peak fluxes measured in $64, 256$ and $1024$ ms bins respectively; duration variables: $T_{50},T_{90}$ are times within which $50\%$ and $90\%$ of the flux arrive.
Unit of fluence: ergs per square centimeter (ergs cm$^{-2}$); unit of peak flux: count per square centimeter per second (cm$^{-2}$ s$^{-1}$); and unit of time: second (s). We compute 
total fluence: $F_T=F_1+F_2+F_3+F_4$ and spectral hardness ratios: $H_{32}=F_3/F_2, H_{321}=F_3/(F_2+F_1)$. Distributions of many variables are significantly correlated and largely skewed, hence we relevantly choose our study variables: $log_{10}T_{50}, log_{10}T_{90}, log_{10}P_{256}, log_{10}F_{T}, log_{10}H_{32}$ and $log_{10}H_{321}$ (Mukherjee et al. 1998; Hakkila et al. 2000; Chattopadhyay et al. 2007, Chattopadhyay and Maitra 2017; Toth et al. 2019). The present analysis include 1956 GRBs possessing finite values on these six variables (Horv\'{a}th et al. 2006.)

\section{Fuzzy clustering}
In clustering the GRBs, we look for meaningful homogeneous groups, where GRBs falling in the existing overlapping clusters can share properties significantly among the groups. In this context, fuzzy clustering methods are quite useful in measuring the extent of how much a GRB is inheriting characteristics from a particular group in terms of membership values, called membership coefficients or we simply refer to them as memberships.  

Here we are trying to cluster $N$ number of GRBs into $K$ fuzzy clusters using FANNY algorithm (Kaufman \& Rousseeuw 2005). It is a nonparametric distance-based machine learning algorithm which is flexible enough to accommodate any distance measure (not necessarily a metric) according to the data. The distance between GRB $i$ and GRB $j$ is denoted by $d(i,j)$ for $i=1,\dots,N$ and $j=1,\dots,N$. In our study, the Euclidean norm is used. Fuzzy clustering provides us with a measure `membership' analogous to probability which describes the degree of certainty that an individual GRB is part of a cluster. Let $m_{ik}$ give the membership of GRB $i$ to cluster $k$, then (1)
	 $m_{ik}\geq0$ for $i=1,\dots,N$ and $k=1,\dots,K$; 
	and (2) $\sum_{k=1}^{K}m_{ik}=1$ for each $i$. Therefore, we can define the membership matrix $M=(m_{ik})_{i=1,\dots,N;k=1,\dots,K}$ associated with clustering of $N$ GRBs into $K$ fuzzy clusters. 

FANNY algorithm is designed to minimize the following objective function:
\begin{equation}
\sum_{k=1}^{K} \frac{\sum_{i=1}^{N}\sum_{j=1}^{N}m_{ik}^rm_{jk}^rd(i,j)}{2\sum_{i=1}^{N}m_{ik}^r},
\end{equation}
where $r$ is a parameter controlling the extent of fuzziness allowed in the algorithm. The value of $r$ is chosen by the analyst depending upon the data, wherein its value needs to be greater than 1 for convergence of the algorithm. In our study, $r=1.3$ is required to carry out fuzzy clustering for $K=3$ and $5$.

In hard clustering, where each GRB is assigned to only one particular cluster, 
$m_{ik}=1$ when GRB $i$ belongs to cluster $k$, otherwise $m_{ik}=0$. This explains for given $K$, $m_{ik}=1$ or $0$ for all $i,k$ result in hard clustering, whereas $m_{ik}=1/K$ for all $i,k$ lead to completely fuzzy clustering. FANNY algorithm also produces the closest hard clustering by placing each GRB to its nearest fuzzy cluster, i.e. the fuzzy class with the highest membership for the burst, wherein larger value of this membership indicates greater certainty in such hard clustering. Proximity between a fuzzy clustering and its nearest hard clustering can be evaluated by the following normalized version of the Dunn's partition coefficient (Dunn 1976; Roubens 1982):
\begin{equation}
N_{DPC}=\frac{(K/N)\sum_{i=1}^{N}\sum_{k=1}^{K}m_{ik}^2-1}{K-1},
\end{equation}
which takes a maximum value of 1 for 
hard clusters and has a minimum value of 0 for a complete fuzzy clustering.

We also plot the non-degenerate principal components (PCs) of the memberships of GRBs in three fuzzy clusters obtained through FANNY algorithm (Rousseeuw et al. 1989), wherein the number of non-degenerate PCs are two since $\sum_{k=1}^{3}m_{ik}=1$ for each $i$. Here principal component analysis is performed on the membership matrix $M$ after each column is standardized (i.e. column-wise mean $=0$ and standard deviation $=1$) so that memberships of the GRBs in three fuzzy clusters are dealt with equal importance (however, non-standardized version also leads to the similar answer).  

\section{Results and interpretation}
\begin{figure}
	\centering
	\includegraphics[width=1\textwidth]{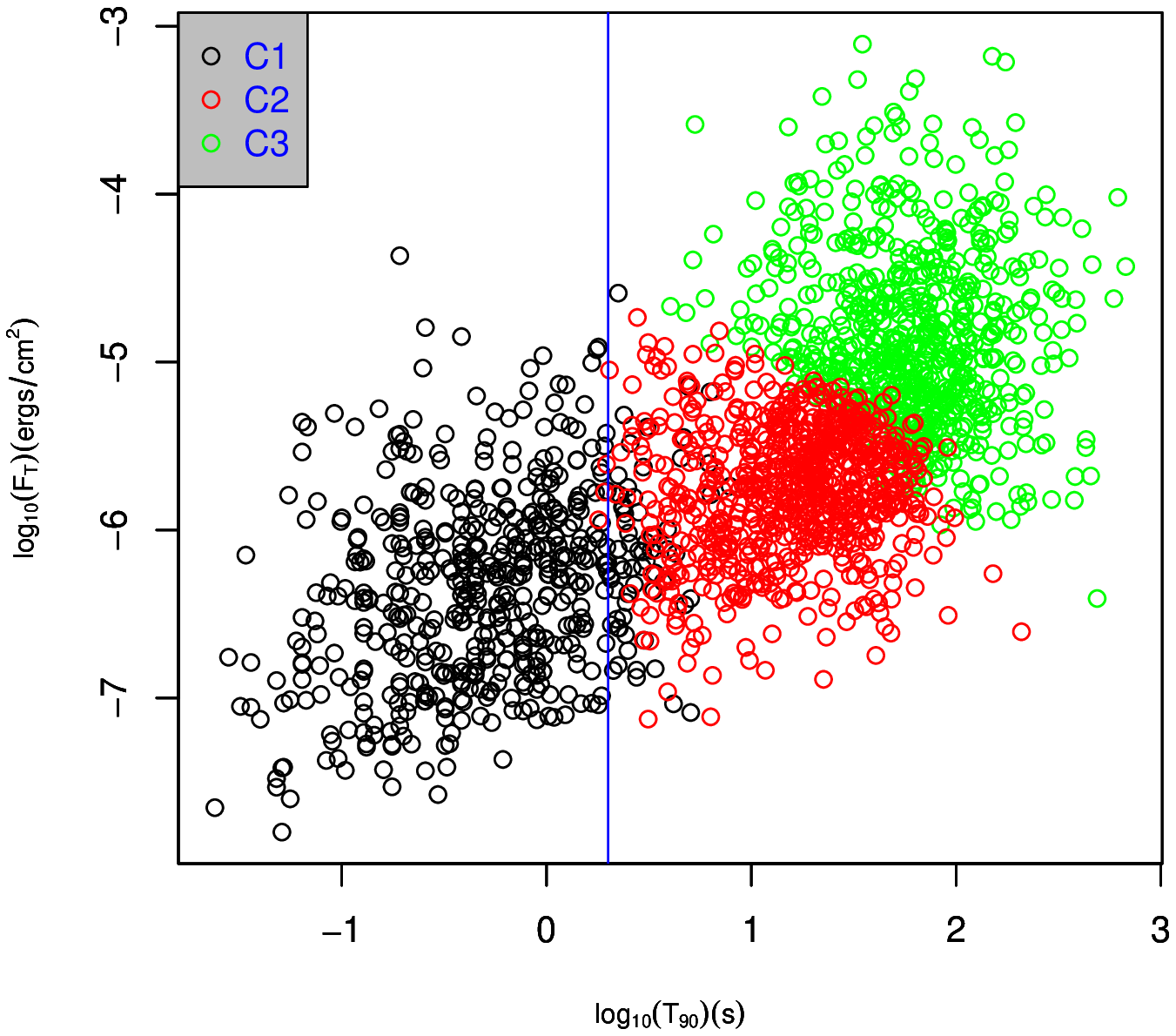}
	\caption{$log_{10}(T_{90})$ (in s) vs. $log_{10}(F_{T})$ (in ergs cm$^{-2}$) plot for three clusters of gamma-ray bursts from the closest hard clustering through FANNY algorithm, wherein the vertical blue line represents $T_{90}=2$ s.}\label{f1}
\end{figure}

\begin{figure}
	\centering
	\includegraphics[width=1\textwidth]{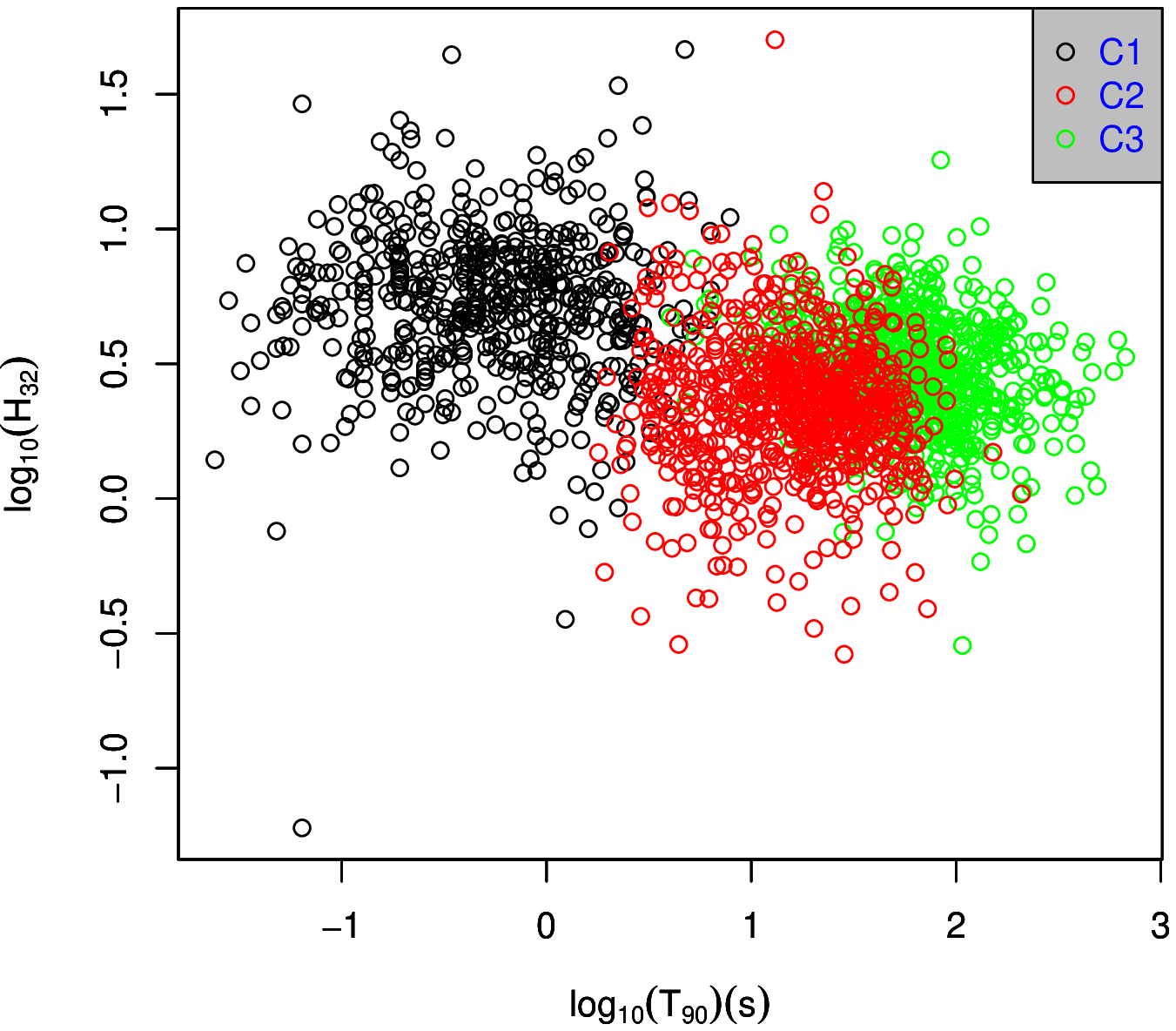}
	\caption{Plot of $log_{10}(T_{90})$ (in s) vs. $log_{10}(H_{32})$ for three clusters of gamma-ray bursts from the closest hard clustering through FANNY algorithm.}\label{f2}
\end{figure}

We carry out fuzzy clustering analysis on the largest GRB data set using FANNY algorithm where we try to distribute the GRBs over three and five clusters, whose closet hard clustering properties are displayed respectively in Tables \ref{t1} and \ref{t2}. Table \ref{t1} prominently exhibits three previously established groups (Mukherjee et al. 1998; Balastegui et al. 2001; Chattopadhyay et al. 2007; Modak et al. 2018), where cluster $C1$ contains the short, dim, hard and lowest--fluence GRBs; $C2$ represents the intermediate, dimmer, soft bursts having lower fluence; whereas cluster $C3$ have the long, brighter, soft bursts with high fluence (see, Figs.~\ref{f1} and \ref{f2}).  

Merely based on the average duration (Table \ref{t2}), in the same fashion as Chattopadhyay \& Maitra (2017, 2018) and Toth et al. (2019), the bursts of five clusters $G1-G5$ can be relatively classified as short, intermediate, intermediate, long and long, respectively. However, in our study, a detailed investigation of the individual GRBs with respect to the well-known three groups $C1-C3$ describes a different scenario. 
Table \ref{t3} reveals clusters $G1$ and $G3$ are built of short and intermediate bursts, respectively; whereas $G4$ consists of the long bursts. Cluster $G5$ is also dominated by and hence corresponds to a group of mainly long bursts but $G2$ is a combination of the short and the intermediate GRBs. 

\begin{figure}
	\centering
	\includegraphics[width=1\textwidth]{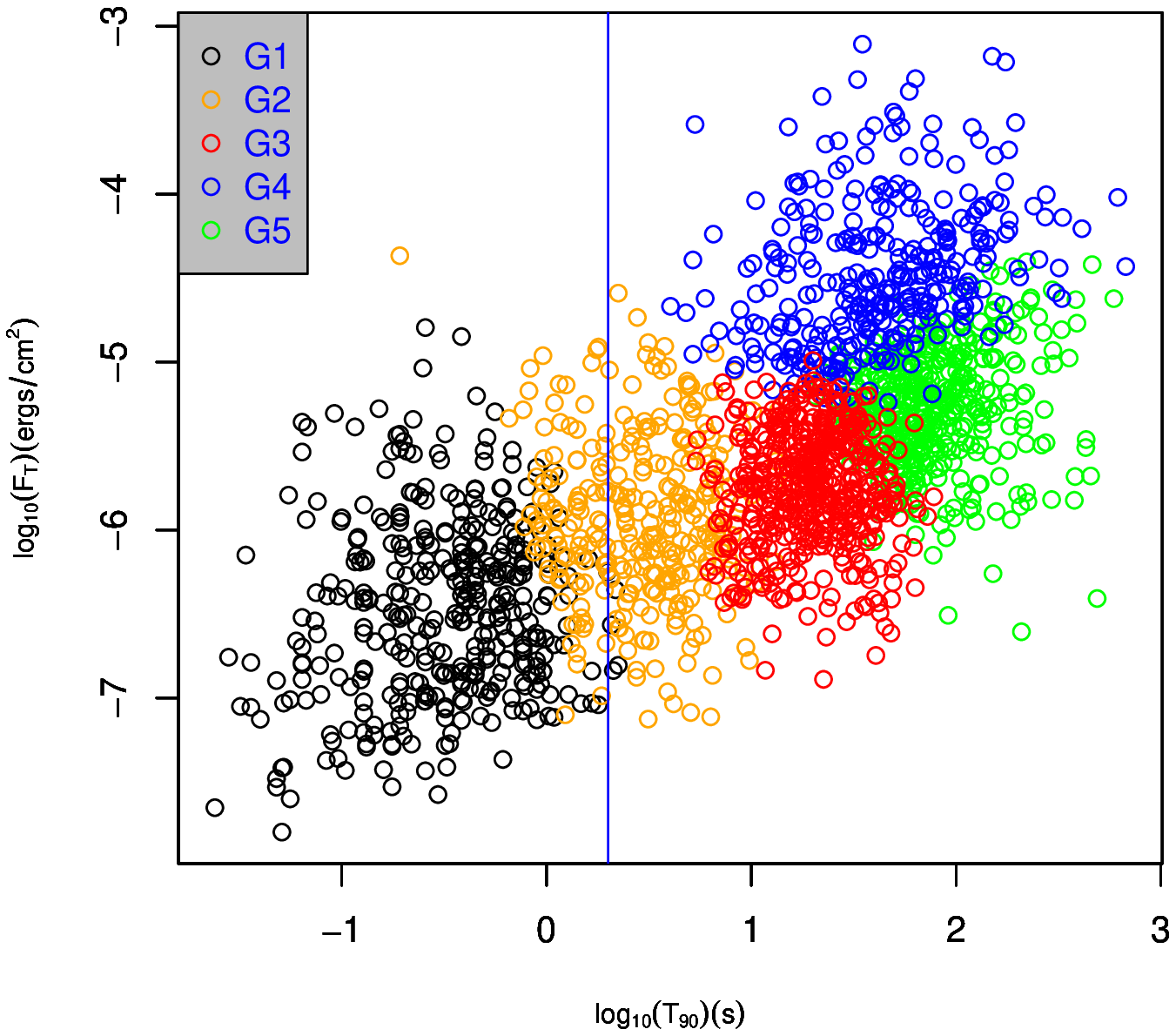}
	\caption{$log_{10}(T_{90})$ (in s) vs. $log_{10}(F_{T})$ (in ergs cm$^{-2}$) plot for five clusters of gamma-ray bursts from the closest hard clustering through FANNY algorithm, wherein the vertical blue line represents $T_{90}=2$ s.}\label{f3}
\end{figure}

\begin{figure}
	\centering
	\includegraphics[width=1\textwidth]{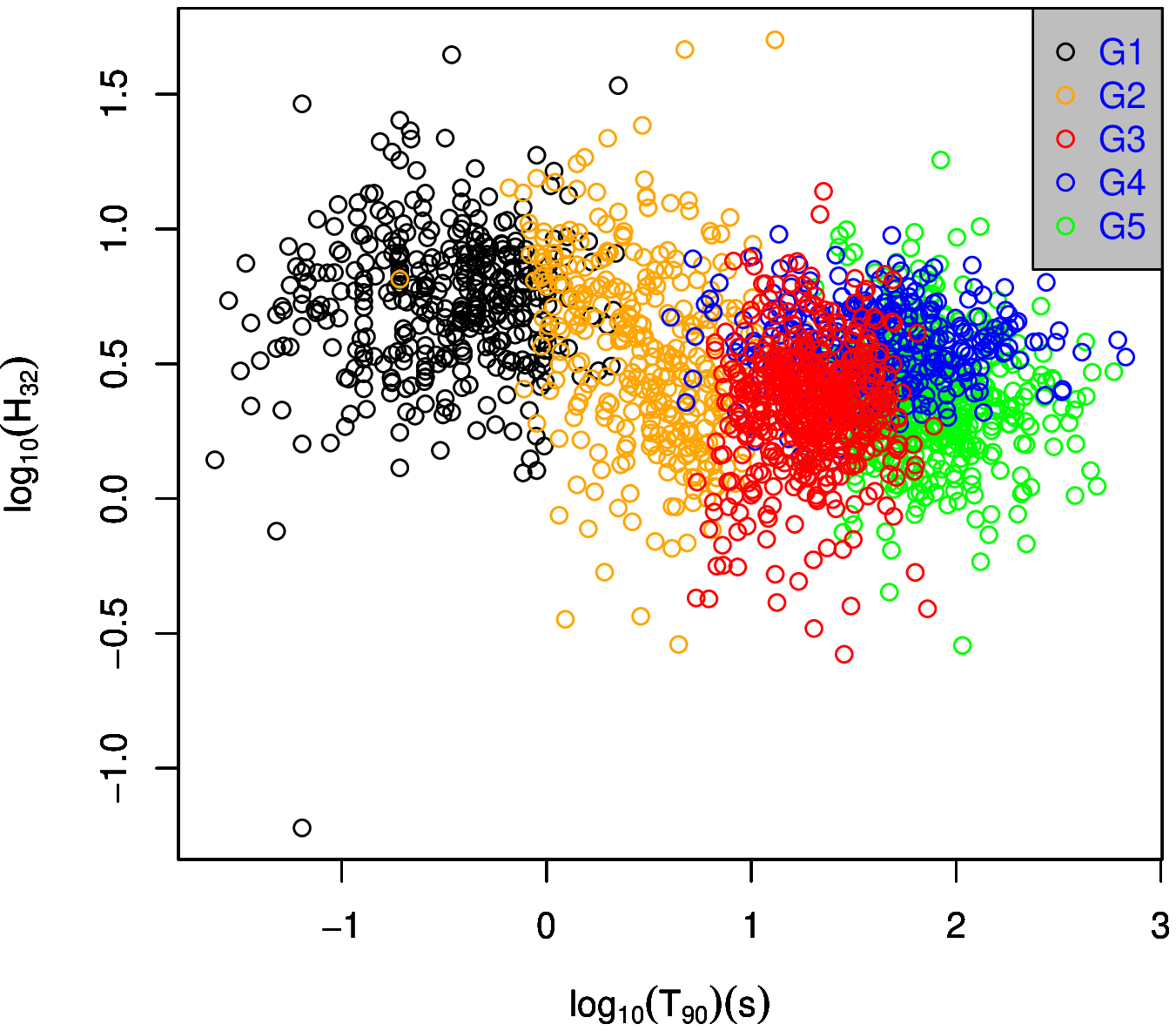}
	\caption{Plot of $log_{10}(T_{90})$ (in s) vs. $log_{10}(H_{32})$ for five clusters of gamma-ray bursts from the closest hard clustering through FANNY algorithm.}\label{f4}
\end{figure}

We investigate whether $G2$ is actually a new group or a spurious cluster. One possibility is that the GRBs, falling near the border of two clusters $C1$ of short bursts and $C2$ of intermediate bursts in three-cluster situation, may now be better clustered in the middle class $G2$ (with average $T_{90}=3.776$ s) lying between $G1$ (with average $T_{90}=0.445$ s) and $G3$ (with average $T_{90}=22.725$ s). In that situation, GRBs from $G2$ should have small memberships in their closest fuzzy cluster $C1$ or $C2$. But we observe that 155 GRBs of $G2$ have a large mean membership of 78.034\% (median $\times 10^2=81.059$) in their nearest fuzzy class $C1$ and the remaining 162 GRBs of $G2$ produce a high mean membership of 70.416\% (median $\times 10^2=70.708$) in their closest fuzzy class $C2$. So in both the cases the high enough membership values prove that all 317 GRBs of $G2$ are well clustered in classes $C1$ and $C2$. Hence the seemingly new cluster $G2$, mixed of short and intermediate bursts, in reality redundantly emerges from numerical separation between clusters $G1$ and $G3$ (see, Figs.~\ref{f3} and \ref{f4}) without any statistical significance. 

\begin{figure}
	\centering
	\includegraphics[width=1\textwidth]{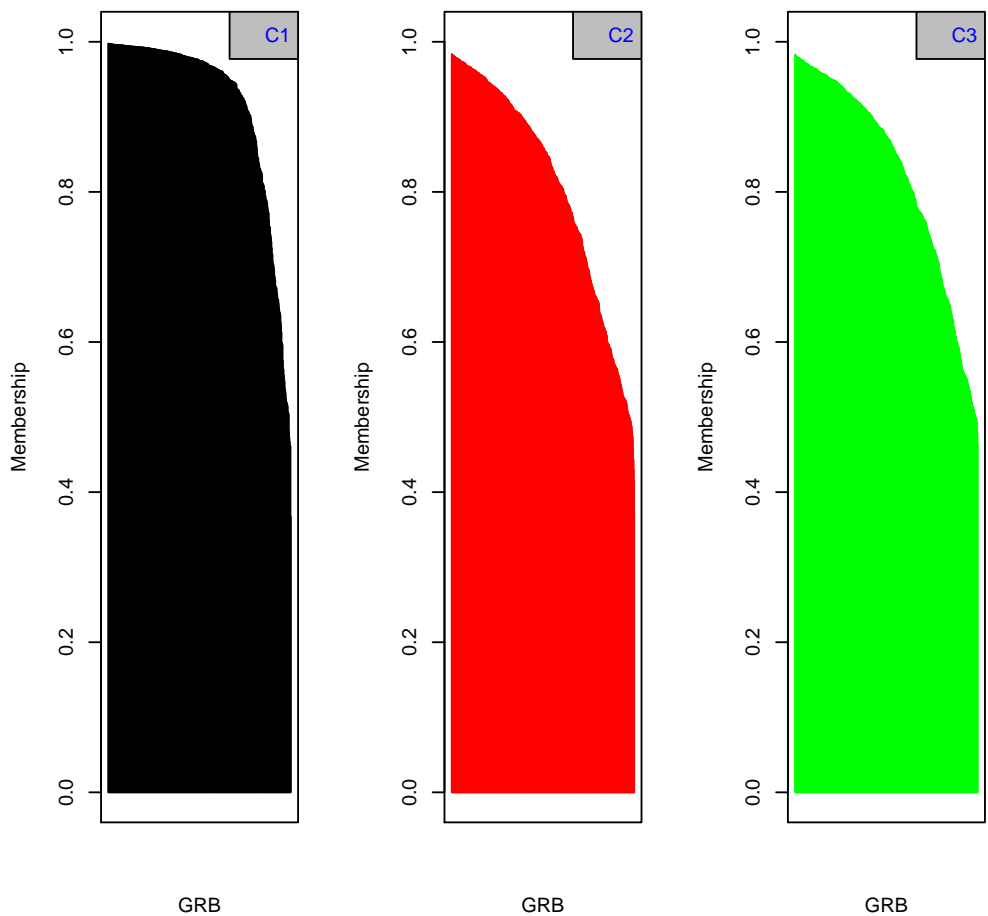}
	\caption{Plot of the memberships of gamma-ray bursts (GRBs) in three fuzzy clusters obtained through FANNY algorithm, where GRBs in their closest fuzzy clusters are arranged monotonically decreasing in membership which is indicated by the length of the vertical shade; from left to right showing classes $C1,C2,C3$ of sizes 529, 742, 685 with respective membership medians $(\times 10^2)$: $97.492, 85.856, 87.394$ and respective membership means 
		(in \%): $91.780,80.822, 82.331$.}\label{f5}
\end{figure}

\begin{figure}
	\centering
	\includegraphics[width=1\textwidth]{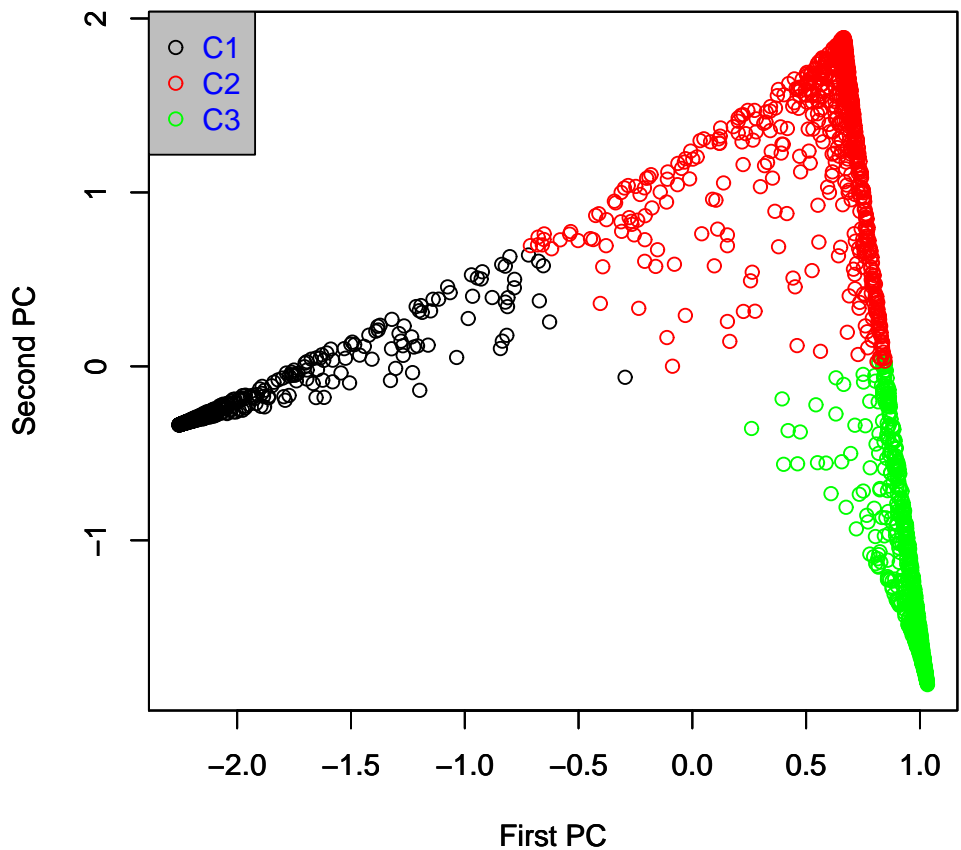}
	\caption{Plot of the two non-degenerate principal components (PCs), explaining 53.65\% and 46.35\% variations, respectively, of the standardized memberships of gamma-ray bursts in three fuzzy clusters obtained through FANNY algorithm; it shows the existence of three distinct hard clusters of gamma-ray bursts.}\label{f6}
\end{figure}

Now, we have a close look at each of the GRBs with respect to membership in its closest fuzzy cluster, i.e. the hard cluster it is assigned to through FANNY algorithm. Classes $C1,C2,C3$ with significantly large memberships, wherein respective medians $(\times 10^2)$ are $97.492, 85.856, 87.394$ and respective means 
(in \%) are $91.780,80.822, 82.331$, and in total only 35 (1.789\%) GRBs have memberships less than 50\%, indicate well clustering of GRBs in the mentioned three clusters (see, Fig.~\ref{f5} for a detailed visual impression of the high memberships in three clusters). Fig.~\ref{f6} clearly shows three distinct clusters of GRBs in terms of the two non-degenerate principal components (PCs) explaining 53.65\% and 46.35\% variations, respectively, of the standardized memberships of GRBs in three fuzzy clusters resulted in FANNY algorithm (explained in Section~3). Therefore, meticulous study on the membership matrix $M$ gives strong evidence in favor of three clusters.

Moreover, FANNY algorithm with $K=3,5$ produces values for $N_{DPC} (\times 10^3$) as $657.068$ and $574.824$, respectively, which state that the closest hard clustering indicated by three fuzzy clusters is way more probable than that by five fuzzy groups. 
Finally, the same is also evidenced by the connectivity index (Handl et al. 2005; Modak et al. 2020), which provides a distance-based hard clustering validity measure having its value between zero and infinity, with a lower value suggests better clustering in terms of tighter groups. Its value for the closest hard clustering obtained through FANNY algorithm with $K=3,5$ comes out respectively as 308.081 and 455.450. Therefore, we conclude GRBs are better clustered in three groups than five, among which $G1$ being part of $C1$ is the short, hard class; spurious cluster $G2$ is combination of short class $C1$ and intermediate class $C2$; $G3$ emerges from intermediate, soft group $C2$; $G4$ and $G5$ are made by separation of long, soft class $C3$ between brighter with high fluence and dimmer with lower fluence groups, respectively (see, Table~\ref{t2}, Figs.~\ref{f3} and \ref{f4}).

Relevantly speaking, our three clusters are consistent with the findings in Modak et al. (2018). Their groups $k1,k2,k3$ are comparable with our present classes $C3,C2,C1$, respectively. Modak et al. (2018) perform $k-$means clustering of the present BATSE GRBs using significant nonlinear features extracted through kernel principal component analysis (Sch\"{o}lkopf \& Smola 2002; Modak et al. 2017). A novel kernel, namely kernel (10) with hyperparameters $s=\sigma_{1}$ and $p=1/2$, gives the best results of three groups based on the first two kernel principal components (KPCs) as indicated by the Dunn index (see, their Table 2). Very obviously, insufficient information in terms of merely the first KPC or irrelevant, noisy features induced by the third KPC may mislead to five clusters. However, efficient statistical machine learning methods like kernel principal component analysis, FANNY algorithm robustly expose the inherent clustering structure in GRB data set. This results in the short, hard, dim, lowest--fluence cluster $C1$; the intermediate, soft, dimmer bursts with lower fluence from class $C2$; and the long, soft, bright GRBs of group $C3$ having high fluence and the highest peak flux (see, Table \ref{t1}, Figs.~\ref{f1} and \ref{f2}). These statistically existing three groups may be generated by different physical processes (Mukherjee et al. 1998; Horv\'{a}th 2002; Chattopadhyay et al. 2007; King et al. 2007; Modak et al. 2018). While the short and the long bursts are usually related to compact binary merger (Paczy\'{n}ski 1986; Usov 1992; Bloom et al. 2006; Nakar 2007; Berger 2014) and massive stellar collapse (Woosley 1993; Paczy\'{n}ski 1998; Woosley \& Bloom 2006; Blanchard et al. 2016), respectively; astrophysical sources behind the intermediate group are yet to be confirmed following future analyses of more data, detailed study of individual GRBs using observed light curve, known redshift, other relevant variables, or fitting appropriate physical models.

\section{Conclusions}
In this paper, we reanalyze the biggest ever data set on GRBs to date from BATSE catalog to verify their natural clustering. Recently model-based clustering methods cast ambiguity by provoking a possibility of the existence of five clusters (Chattopadhyay \& Maitra 2017, 2018; Toth et al. 2019) against the three well-known groups composed of short, intermediate and long bursts (Mukherjee et al. 1998; Balastegui et al. 2001; Chattopadhyay et al. 2007; Modak et al. 2018). However, the additional classes are so far at a loss to explain any new astrophysical sources, whereas Toth et al. (2019) study their cluster-wise average properties and conclude that these extra classes are further split-ups of the previously established three groups.

Based on the analysis of duration, hardness and peak flux variables, Toth et al. (2019) describe the new five clusters as short group, division of both intermediate and long groups into two further classes as dimmer and brighter (see, their Table~1). Again, study on the duration, fluence, hardness and peak flux variables results in separation of both the short and the intermediate clusters into dimmer and brighter ones, and one long cluster (see, their Table~2). It is worth noting that such decisions are made by studying group-wise average values of the variables rather than exploring the individual GRBs. They dismiss the statistically found five clusters as a result caused by violation of parametric distributional model assumptions adopted
by the clustering methods. Hence, we apply the robust, nonparametric fuzzy clustering method using FANNY algorithm (Kaufman \& Rousseeuw 2005), where each GRB is studied meticulously by means of its memberships in the fuzzy clusters. We statistically show that the five groups redundantly emerge from numerical separation of the established three groups and therefore solve the new conflict with confirmation of existence of three previously known clusters in GRBs whose astrophysical sources are still undergoing extensive investigation.
\section{Acknowledgments}

The author would like to thank the Editor and two anonymous referees for encouraging the present work and is sincerely grateful to the referees for their intriguing inquiries and insightful comments which helped to present the manuscript in a more precise way.
\clearpage
\begin{table}
	\caption{Properties of three groups (mean value with standard error) from the closest hard clustering through FANNY algorithm}
	\begin{center}
		\tiny
		\begin{tabular}{c c c c c c c c}
			\hline\\
Cluster&Cluster-size & $T_{50}$ &  $T_{90}$ & $P_{256}$ & $F_{T} \times 10^{6}$ & $H_{32}$ &$H_{321} $\\
name &  (percentage)        & (s)     & (s)     & (cm$^{-2}$ s$^{-1}$) &  (ergs cm$^{-2}$)      & &\\[1ex]
			\hline\\
			$C1$       & 529 (27.044\%)  & 00.327 $\pm$ 0.015  & 00.921 $\pm$ 0.045 & 2.363 $\pm$ 0.160
			&01.187 $\pm$  0.121 & 6.425 $\pm$ 0.206 &  3.937 $\pm$ 0.101\\[1ex]
			
			$C2$       & 742 (37.934\%)  &  08.141 $\pm$ 0.297 & 21.979 $\pm$ 0.677& 1.598 $\pm$ 0.087 
			&02.467 $\pm$ 0.079 & 2.681 $\pm$ 0.089 & 1.471 $\pm$ 0.038\\[1ex]
			
			$C3$       & 685 (35.020\%) & 37.759 $\pm$ 2.039 & 85.361 $\pm$ 3.082& 5.700 $\pm$ 0.463
			& 32.639 $\pm$ 2.636 & 3.310 $\pm$ 0.065 & 1.941 $\pm$ 0.042\\[1ex]
			\hline
		\end{tabular}
	\end{center}
	\label{t1}
\end{table}
\clearpage
\begin{table}
	\caption{Properties of five groups (mean value with standard error) from the closest hard clustering through FANNY algorithm}
	\begin{center}
		\tiny
		\begin{tabular}{c c c c c c c c}
			\hline\\
			Cluster&Cluster-size & $T_{50}$ &  $T_{90}$ & $P_{256}$ & $F_{T} \times 10^{6}$ & $H_{32}$ &$H_{321} $\\
		name	&       (percentage)   & (s)     & (s)     & (cm$^{-2}$ s$^{-1}$) &  (ergs cm$^{-2}$)  & &\\[1ex]
			\hline\\
			$G1$       & 374 (19.120\%) & 00.167 $\pm$ 0.007  & 00.446 $\pm$ 0.019 & 02.094 $\pm$ 0.142
			&00.763 $\pm$  0.076 & 6.417 $\pm$ 0.235 &  4.022 $\pm$ 0.120\\[1ex]
			
			$G2$       & 317 (16.206\%) & 01.265 $\pm$ 0.044 & 03.776 $\pm$  0.151& 02.992 $\pm$ 0.266 
			& 02.325 $\pm$ 0.211 & 4.863 $\pm$ 0.282 & 2.727 $\pm$  0.120\\[1ex]
			
			$G3$       & 501 (25.614\%) & 08.104 $\pm$ 0.225 & 22.726 $\pm$ 0.539&  01.182 $\pm$ 0.041
			& 02.337 $\pm$ 0.075 & 2.513 $\pm$ 0.064 & 1.396 $\pm$ 0.040\\[1ex]
			
			$G4$      &  327 (16.718\%)  & 22.089 $\pm$ 1.918 & 64.438 $\pm$ 4.142& 10.625 $\pm$ 0.895
			& 59.709 $\pm$ 5.112 & 3.876 $\pm$ 0.075 &  2.355 $\pm$ 0.055\\[1ex]
			
			$G5$      & 437 (22.342\%)  & 46.526 $\pm$ 2.786 & 94.846 $\pm$ 3.693& 01.241 $\pm$ 0.040
			& 07.091 $\pm$  0.278 & 2.718 $\pm$ 0.083 & 1.522 $\pm$ 0.051\\[1ex]
			\hline
		\end{tabular}
	\end{center}
	\label{t2}
\end{table}
\clearpage
\begin{table}
	\caption{Number of gamma-ray bursts (\%) in three and five groups from the closest hard clusterings obtained through FANNY algorithm}
	\begin{center}
		\tiny
		\begin{tabular}{c| c c c|c}
\hline\\ Cluster & $C1$               &    $C2$         &    $C3$        & Total\\[1ex]
\hline\\
         $G1$    & 374 (100\%)     &0             &    0        & 374 (100\%)\\[1ex]
         $G2$    & 155 (48.896 \%) & 162 (51.104\%)&0            & 317 (100\%)\\[1ex]
         $G3$   & 0               &492 (98.203\%) &9 (1.796\%)   & 501 (100\%)\\[1ex]
         $G4$    & 0               &11 (3.364\%)  &316 (96.636\%)&327 (100\%)\\[1ex]
         $G5$     & 0               &77 (17.620\%)  &360 (82.380\%)&437 (100\%)\\[1ex]
         
			\hline
	
		\end{tabular}
	\end{center}
	\label{t3}
\end{table}
\clearpage
\clearpage
\clearpage
\clearpage
\clearpage
\clearpage

\end{document}